\documentstyle[psfig]{mn}

\title[The binary fraction of EHB stars.]
{The binary fraction of extreme horizontal branch  stars.}
\author[P. F. L. Maxted et~al.]
       {P. F. L. Maxted$^{1,2}$, U. Heber$^3$,  T. R. Marsh$^1$, 
 R.C. North$^1$ \\
 $^1$University of Southampton, Department of Physics \& Astronomy,
 Highfield, Southampton, S017 1BJ, UK \\
 $^2$ Department of Physics, Keele University, Staffordshire, ST5~5BG,
 UK\\
 $^3$ Dr. Remeis-Sternwarte, Astronomisches Institut der Universit\"{a}t
      Erlangen-N\"{u}rnberg, Sternwartstrasse 7, 96049 Bamberg, Germany}
\date{Accepted 2000
      Received 2000 }

\pagerange{\pageref{firstpage}--\pageref{lastpage}}
\pubyear{2000}

\newcommand{\Msolar}{\mbox{${\rm M}_{\odot}$}}
\newcommand{\Rsolar}{\mbox{${\rm R}_{\odot}$}}
\begin{document}

\maketitle

\label{firstpage}
\begin{abstract}
 We have used precise radial velocity measurements of subdwarf-B stars
from the Palomar-Green catalogue to look for binary extreme horizontal branch
(EHB) stars. We have determined the effective temperature,  surface gravity
and surface helium abundance for 20 of the targets from new or existing
blue spectra and have compiled published values for these quantities for all
but one other. We identify 36 EHB stars in our sample and find that at least
21 of these stars are binaries. All but one or two of these are new
identifications. The minimum binary fraction for EHB stars implied by our
survey is 60$\pm$8\,percent. Our survey is sensitive to binaries with orbital
periods $P\la 10$\,d. For reasonable assumptions concerning the period
distribution and the mass ratio distribution of the binaries, we find that the
mean detection efficiency of our survey over this range of orbital periods is
87\,percent. Allowing for this estimated detection efficiency, the fraction of
EHB stars which are short-period binaries  ($0.03{\rm \,d} \la P\la 10$\,d) is
69$\pm$9\,percent. The value is not strongly dependent on the period
distribution below $P\approx 10$\,d or the mean companion mass for these
short-period binaries. The orbital separation of the stars in these binaries
is much less than the size of the red giant from which the EHB star has
formed. This is strong evidence that binary star evolution is fundamental to
the formation of the majority of EHB stars. If there are also binary EHB stars
whose orbital periods are $\ga 10$\,d, the fraction of EHB stars whose
evolution has been affected by the presence of a companion may be much higher,
e.g., if 1/3 of EHB stars are binaries with orbital periods 10\,d$\la P \la
100$\,d, then our observations are consistent with all EHB stars being formed
through some type of binary star evolution.  We find that 5 of the other stars
we observed are likely to be post-EHB stars, one of which is also a binary. 
\end{abstract}


\section{Introduction}
 Surveys for  blue stars brighter than $B\approx 16$ are dominated by
subdwarf-B (sdB) stars (Green, Schmidt \& Liebert 1986). The effective
temperatures (T$_{\rm eff}$) and surface gravities ($\log g$) of the majority
of these stars place them on the extreme horizontal branch (EHB), i.e., they
appear in the  same region of the T$_{\rm eff}$\,--\,$\log g$ plane as
evolutionary tracks for core helium burning stars with core masses of about
0.5\Msolar\ and extremely thin ($\la 0.02\Msolar$) hydrogen envelopes (Heber
1986; Saffer et~al. 1994). We make a distinction in this paper between the
nomenclature ``sdB star'', which is  a spectral classification, and ``EHB
star'' which is an interpretation of the evolutionary state of a star.

 The observed dispersion of core masses for EHB stars is very low ($<0.04
\Msolar$, Saffer et~al. 1994). It is thought that the eventual fate of an EHB
star is to cool to form a white dwarf with a mass of about 0.5\Msolar, which
is low compared to the typical mass for white dwarfs (Bergeron, Saffer \&
Liebert 1992). The formation of low mass white dwarfs is, in general, thought
to involve interactions with a binary companion star, e.g., a common envelope
phase, in which a companion to a red giant star is engulfed by the expanding
outer layers. The resulting friction causes the companion to spiral  in
towards the core of the red giant, ejecting the envelope at the expense of
orbital binding energy (Iben \& Livio 1993). If this process occurs while the
red giant is within $\sim$0.4\,magnitudes of the tip of the red giant branch,
the core can go on to ignite helium, despite the dramatic mass loss, and may
then appear as an EHB star (D'Cruz et~al. 1996; Mengel, Norris \&  Gross
1976). 

 The binary fraction of sdB and EHB stars is expected to be high given the
scenario outlined above. Allard et~al. (1994) found that 31 of their sample of
100 sdB stars show flat spectral energy distributions which indicate the
presence of companions with spectral types in the range late-G to early-M.
They infer a binary fraction for main-sequence companions of
54\,--\,66\,percent, although the companions in their survey appear to be
over-luminous compared to normal main-sequence stars. A similar conclusion was
reached by Ferguson, Green \& Liebert (1984) using a similar argument and by
Jeffery \& Pollacco (1998) based on the detection of spectral features due to
cool companions. What is not clear from these observations is whether the cool
companion is sufficiently close to the EHB star to be implicated in the mass
loss process that is supposed to form the EHB star. These techniques are also
insensitive to white dwarf companions and faint M-dwarf companions. Companions
to EHB stars can also be detected in eclipsing systems such as the
short-period EHB\,--\,M-dwarf binaries HW~Vir (Wood \& Saffer 1999) and
PG\,1336$-$018 (Kilkenny et~al. 1998) and in EHB\,--\,white dwarf binaries
which show ellipsoidal variability, e.g.,  KPD\,0422+5421 (Koen et~al. 1998)
and KPD\,1930+2752 (Maxted, Marsh \& North 2000). These binaries are extremely useful
for studying the properties of EHB stars, but they  do not offer a useful
method for finding binary EHB stars in general because the probability of such
a binary showing eclipses or a measureable ellipsoidal effect decreases
rapidly for increasing orbital periods.

Radial velocity surveys are an excellent method for identifying binary stars in
general, particularly since the effeciency of this technique can be accurately
quantified and the selection effects are well understood. This technique can
be applied to many types of star, from  main-sequence stars (Duquennoy \&
Mayor 1991) to white dwarfs (Maxted \& Marsh 1999).  For EHB stars in
particular, this is the method of choice because the short-period binaries
which are expected to result from a common-envelope phase are the easiest to
identify using this method. If the radial velocities are measured to an
precision of a few km\,s$^{-1}$ and the observations are obtained over a
baseline of weeks or months the technique has the potential to identify
binaries with much longer periods ($\sim$ 100\,d) even if the companion is a
low mass M-dwarf. Saffer, Livio \& Yungelson (1998) have shown  the potential
for this technique with their observations of 46 sdB stars. The precision of
their radial velocity  measurements was modest (20--30\,km\,s$^{-1}$) and the
three spectra they obtained for each star over a baseline of 1--2 days were
compared by-eye, yet they found that at least 7 of their sample of 46 sdB
stars show radial velocity variations. Several of these binaries have
subsequently had their orbital periods determined (Moran et~al. 1999, Maxted
et~al. 2000), although further observations are required to determine the
nature of the companions in these binaries.  

 In this paper we present the results of a radial velocity survey of binary
EHB stars. We have used observations of the H${\alpha}$ line to measure 205
precise  radial velocities  for 36 EHB stars from the Palomar-Green survey
(Green, Schmidt \& Liebert 1986). We positively identify 22 short-period
binary EHB stars, 20 of which are new discoveries. We conclude that at least
60$\pm$8\,percent of of EHB stars are short-period binary stars. If we allow
for the detection efficiency of our survey, we find that at least
69$\pm$9\,percent of EHB stars are binaries. This is strong evidence that
binary star evolution is fundamental to the formation of the majority of EHB
stars.  We also observed 5 stars which we identify as post-EHB stars and found
that one of these stars is a binary.

\section{Observations and reductions}
\subsection{H$\alpha$ spectra}
 Targets were selected from objects in the Palomar-Green catalog (Green,
Schmidt \& Liebert 1986) classified as sdB stars. We avoided stars where
follow-up observations have shown the classification was in error or that the
star is not an EHB star. Observations were obtained with the 2.5m Isaac Newton
Telescope on the Island of La Palma. Spectra were obtained with the
intermediate dispersion spectrograph  using the 500mm camera, a 1200~line/mm
grating and a TEK charge coupled device (CCD) as a detector. The spectra cover
400\AA\ around the H${\alpha}$ line at a dispersion of 0.39\AA\ per pixel. The
slit width used was 0.97\,arcsec which gave a resolution of about 0.9\AA.
Spectra of the targets were generally obtained in pairs bracketed by
observations of a copper-neon arc. We obtained a total of 243 spectra for 43
stars over a total of about 7 nights during the interval 2000 April
10\,--\,21. The seeing was good ($\approx$ 1\,arcsec) on most of these nights.

  Extraction of the spectra from the images was performed automatically using
optimal extraction to maximize the signal-to-noise of the resulting spectra
(Marsh 1989). The arcs associated with each stellar spectrum were extracted
using the same weighting determined for the stellar image to avoid possible
systematic errors due to the tilt of the spectra on the detector.  The
wavelength scale was determined from a fourth-order polynomial fit to measured
arc line positions. The standard deviation of the fit to the 8 arc lines
was typically 0.09\AA. The wavelength scale for an individual spectrum was
determined by interpolation to the time of mid-exposure from the fits to arcs
taken before and after the spectrum to account for the small amount of drift
in the wavelength scale ($<0.1$\AA) due to flexure of the instrument.
Statistical errors on every data point calculated from photon statistics are
rigorously propagated through every stage of the data reduction.

\subsection{Blue spectra}

 In order to measure the effective temperature and surface gravity of some our
targets we also obtained blue spectra of our targets with the same telescope
and instrument. We did not attempt to measure radial velocities from these
spectra.

Spectra of PG\,1032+406, PG\,1043+760, PG\,1051+501, PG\,1039+219,
PG\,1043+760 and  PG\,1110+294 were obtained over the wavelength range
3810--5020\AA\ using a 400~line/mm grating. The observations were obtained
while the stars were at low airmass in good seeing with a vertical 1.5\,arcsec
slit. The resolution in pixels was determined from the width of the spatial
profile, which was typically 3\,--\,4 pixels which corresponds to a resolution
of about 4\AA.

 We used an EEV CCD on the 235mm camera and a 900~line/mm grating to obtain
spectra of PG\,1505+074, PG\,1512+244 and PG\,1553+273 on the night 2000 July
16 and of PG\,1616+144, PG\,1627+017, PG\,1632+088, PG\,1647+056, and
PG\,1653+131 on the night 2000 August 15. The useful region of the spectra
cover the wavelength range 3590\,--\,5365\AA\ at a dispersion of 0.63\AA\ per
pixel. We used a vertical, 1\,arcsec wide slit which  gave a resolution of
1.6\AA. We also obtained spectra of PG\,0907+123 and PG\,1116+301 on the night
of 2001 February 3 with the same instrument covering the wavelength range
3850\,--\,5200\AA. One other spectrum of PG\,0907+123 was also obtained and
reduced for us by Martin Altman using the Calar Alto 2.2m telescope and the
CAFOS spectrograph with a B100 grism at a dispersion of 100\AA/MM at lower
spectral resolution than our INT spectra, but covering the Balmer lines from
H$\beta$ to H10.

\section{Analysis}

\subsection{\label{TeffLogg} Effective temperatures, surface  gravities and
helium abundances.}

 For those stars for which we have blue spectra we measured the effective
temperature, T$_{\rm eff}$, the surface  gravity $\log g$ and the helium
abundance by number, $y$. The simultaneous fitting of Balmer line profiles by
a grid of synthetic spectra has become the standard technique to determine the
atmospheric parameters of hot high gravity stars (Bergeron et al. 1992). The
procedure has been extended to include helium line profiles and
applied successfully to sdB stars by Saffer et al. (1994). We have applied
Saffer's procedure to the Balmer lines (H$\beta$ to H$\,$9), and the He~I
(4026\AA, 4388\AA, 4471\AA, 4713\AA, 4922\AA) and He~II 4686\AA\ lines.

\begin{figure*}
\psfig{file=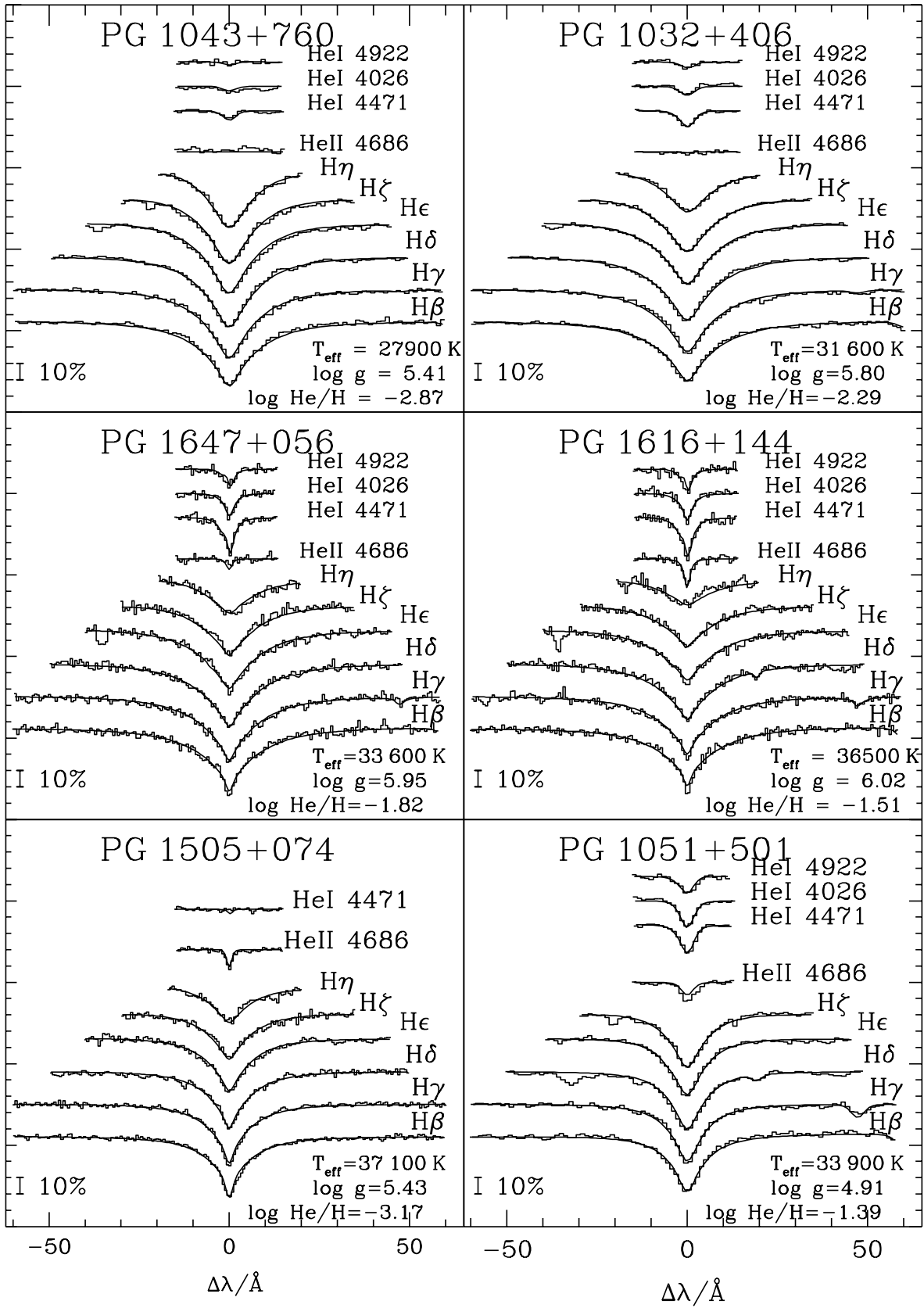,width=0.95\textwidth} 
\caption{\label{BlueFitFig}Examples of the observed spectral lines and
synthetic spectrum fits for six sdB stars.}
\end{figure*}

A grid of synthetic spectra derived from H and He line blanketed NLTE 
model atmospheres (Napiwotzki 1997) was matched to the data to simultaneously
determine the effective temperature, surface gravity and helium abundance. For
stars cooler than 27\,000\,K we used the metal line-blanketed LTE model
atmospheres of Heber, Reid, \& Werner (2000). The synthetic spectra were
convolved beforehand with a Gaussian profile of the appropriate width to
account for the instrumental profile.  The adopted values of T$_{\rm eff}$,
$\log g$ and $y$ for these stars and all other stars where values could be
found are given in Table~\ref{TeffloggTable}. Examples of the observed
spectral lines and synthetic spectrum fits for six sdB stars are shown in
Fig.~\ref{BlueFitFig}. The values given for PG\,1040+234 are only approximate
because of the contamination by the companion star, particularly in the
H$\epsilon$ line, which we excluded from the fit. The values given for
PG\,1701+359, PG\,1722+286 and PG\,1743+477 are based on updated fits to the
spectra described Theissen et~al. (1993). The values for PG\,0907+123 are an
average of the values derived from our INT spectrum and the spectrum taken by
Martin Altman, which agree very well.

\begin{figure}
\psfig{file=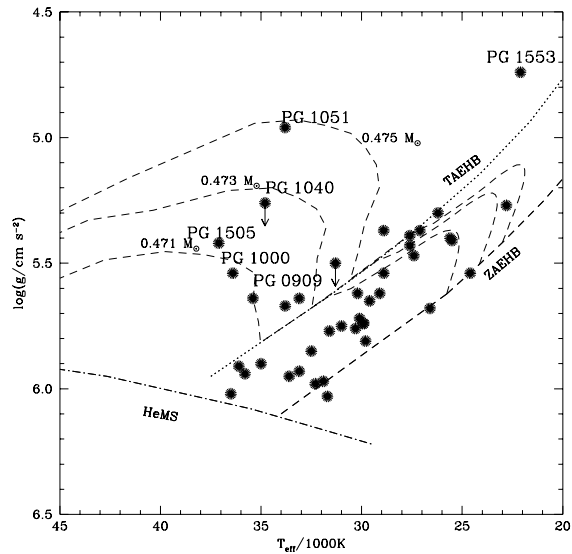,width=0.45\textwidth} 
\caption{\label{TeffloggFig} The measured values of T$_{\rm eff}$ and $\log g$
for our targets compared to the evolutionary models of Dorman et~al. (1993).
Evolutionary tracks are shown as dashed lines and  are labelled with the mass
of the helium core they refer to. The other symbols are defined in the text.
Stars discussed in the text are labelled by the first 6 characters of their PG
catalogue names. Small arrows denote measurements which are upper limits.}
\end{figure}

 The measured values of  T$_{\rm eff}$ and $\log g$ are compared to the
evolutionary tracks for extreme horizontal branch stars of Dorman et~al.
(1993) in Fig.~\ref{TeffloggFig}. It can be seen that most the targets lie in
or near the band defined by the zero-age extreme horizontal branch (ZAEHB),
the terminal-age extreme horizontal branch (TAEHB) and the helium
main-sequence (HeMS) and are therefore EHB stars. The errors in the
atmospheric parameters are estimated to be 3\% for T$_{\rm eff}$ and
0.1 dex for $\log g$ and the helium abundance $y$. We consider a star to be a
post-EHB star if it lies above the TAEHB by more than these error bars,
otherwise we regard it as an EHB star. The spectra of PG\,1040+234 and
PG\,1701+359 are contaminated by the cool companion. The principal effect of
this contamination is to bias the value of $\log g$ to lower values,  so we
treat the value obtained as an upper limit in Table~\ref{TeffloggTable} and in
Fig.~\ref{TeffloggFig}. Both these stars are included as EHB stars in our
discussion of the binary fraction of EHB stars. PG\,1632+088 is too cool to
appear in Fig.~\ref{TeffloggFig}. It is probably a normal horizontal branch
star so we exclude it from our discussion of the binary fraction of EHB stars.
We also exclude  PG\,0909+164, PG\,1000+408, PG\,1051+501, PG\,1505+074 and
PG\,1553+273 because they lie too far from the EHB in the T$_{\rm
eff}$\,--\,$\log g$ plane. They appear to be more evolved than EHB stars so we
classify them as post-EHB stars and discuss them separately from the EHB
stars. We also exclude PG\,1631+267 from our discussion of the binary fraction
of EHB stars because no T$_{\rm eff}$\,--\,$\log g$ measurement is available
for this star.

\begin{table}
\caption{\label{TeffloggTable}
 Measured values of T$_{\rm eff}$, $\log g$ and $y$ from blue spectra for our
targets. We also give the $y$ magnitude  in the Str\"omgren system, $m_y$, from 
Wesemael et al. (1992) and Bergeron et al. (1984). 
References are as follows: 0.~This work; 1.~Saffer et~al. (1994);
2. Saffer, priv. comm.; 3.~Moehler et~al. (1990); 4.~O'Donoghue et~al.(1998).
}
\begin{tabular}{@{}lrrrrlc}
Name &  $m_y$ & \multicolumn{1}{l}{T$_{\rm eff}$}
     &\multicolumn{1}{l}{$\log g$ }  
     &\multicolumn{1}{c}{$y$}& Ref. \\
     &   & \multicolumn{1}{l}{(kK)}
     &\multicolumn{1}{l}{(cgs)} \\
 PG\,0749+658   & 12.14 & 24.6  & 5.54  & 0.004   & 1  \\
 PG\,0839+399   & 14.39 & 36.1  & 5.91  & 0.002   & 1  \\
 PG\,0849+319   & 14.61 & 28.9  & 5.37  & 0.003   & 2  \\
 PG\,0850+170   & 13.98 & 27.1  & 5.37  & 0.006   & 2  \\
 PG\,0907+123   & 13.97 & 26.2  & 5.30  & 0.018   & 0  \\
 PG\,0909+164  & 13.85 & 35.4  & 5.64  & 0.002   & 2  \\
 PG\,0918+029   & 13.42 & 31.7  & 6.03  & 0.008   & 1  \\
 PG\,0919+273   & 12.77 & 31.9  & 5.97  & 0.011   & 1  \\
 PG\,1000+408  & 13.33 & 36.4  & 5.54  & 0.002   & 2  \\
 PG\,1017$-$086 & 14.43 & 30.2  & 5.62  & 0.003   & 2  \\
 PG\,1018$-$047 & 13.32 & 31.0  & 5.75  & 0.002   & 2  \\
 PG\,1032+406   & 11.52 & 31.6  & 5.77  & 0.005   & 0  \\
 PG\,1039+219   & 13.09 & 33.1  & 5.64  & 0.007   & 0  \\
 PG\,1040+234$^a$  & 13.37 & 34.8  &$>5.26$&$>0.030$ & 0 \\
 PG\,1043+760   & 13.77 & 27.6  & 5.39  & 0.002   & 0  \\
 PG\,1047+003   & 13.48 & 35.0  & 5.9   &         & 4  \\
 PG\,1051+501  & 13.38 & 33.8  & 4.96  & 0.040   & 0  \\
 PG\,1110+294   & 14.09 & 30.1  & 5.72  & 0.019   & 0  \\
 PG\,1114+073   & 13.06 & 29.8  & 5.81  & 0.006   & 1  \\
 PG\,1116+301   & 14.34 & 32.5  & 5.85  & 0.006   &  0  \\
 PG\,1237+132   & 14.65 & 33.1  & 5.93  & 0.002   & 2  \\
 PG\,1244+113   & 14.20 & 33.8  & 5.67  & 0.001   & 2  \\
 PG\,1248+164   & 14.40\makebox[0pt][l]{$^b$} & 26.6  & 5.68  & 0.001   & 2 \\
 PG\,1300+279   & 14.27 & 29.6  & 5.65  & 0.005   & 2  \\
 PG\,1303+097   & 14.50 & 30.3  & 5.76  & 0.011   & 2  \\
 PG\,1329+159   & 13.55 & 29.1  & 5.62  & 0.004   & 2  \\
 PG\,1417+257   & 13.78 & 27.6  & 5.43  & 0.005   & 2  \\
 PG\,1505+074  & 12.44 & 37.1  & 5.42  & 0.0008  & 0  \\
 PG\,1512+244   & 13.28 & 29.9  & 5.74  & 0.009   & 0  \\
 PG\,1553+273   & 13.61 & 22.1  & 4.74  & 0.001   & 0  \\
 PG\,1616+144   & 13.50 & 36.5  & 6.02  & 0.031   & 0  \\
 PG\,1619+522   & 13.30 & 32.3  & 5.98  & 0.011   & 1  \\
 PG\,1627+017   & 12.93 & 22.8  & 5.27  & 0.001   & 0  \\
 PG\,1631+267   & 15.51 &       &       &         & -  \\
 PG\,1632+088   & 13.19 & 13.3  & 3.78  & 0.004   & 0  \\
 PG\,1647+056   & 14.75 & 33.6  & 5.95  & 0.015   & 0  \\
 PG\,1653+131   & 14.50 & 25.6  & 5.40  & 0.002   & 0  \\
 PG\,1701+359$^a$ & 13.22 & 31.4  &$>$5.50  &$>$0.0003  & 0  \\
 PG\,1710+490   & 12.90 & 29.9  & 5.74  & 0.006   & 1  \\
 PG\,1716+426   & 13.97 & 27.4  & 5.47  & 0.003   & 1  \\
 PG\,1722+286   & 13.40 & 35.8  & 5.94  & 0.035   & 0  \\
 PG\,1725+252   & 13.01 & 28.9  & 5.54  & 0.0009  & 0  \\
 PG\,1743+477   & 13.79 & 25.5  & 5.41  & 0.007   & 0  \\
\hline
\multicolumn{6}{l}{$^a$ spectrum contaminated by cool companion}\\
\multicolumn{6}{l}{$^b$ photographic magnitude}\\
\end{tabular}
\end{table}

\subsection{Radial velocity measurements.}
  To measure the radial velocities we used least-squares fitting of a
model line profile. This model line profile is the summation of three Gaussian
profiles with different widths and depths but with a common central position
which varies between spectra.  Only data within 2000\,km\,s$^{-1}$ of the
H${\alpha}$ line is included in the fitting process and the spectra are
normalized  using a linear fit to the continuum either side of the H${\alpha}$
line. 
 
\begin{figure}
\psfig{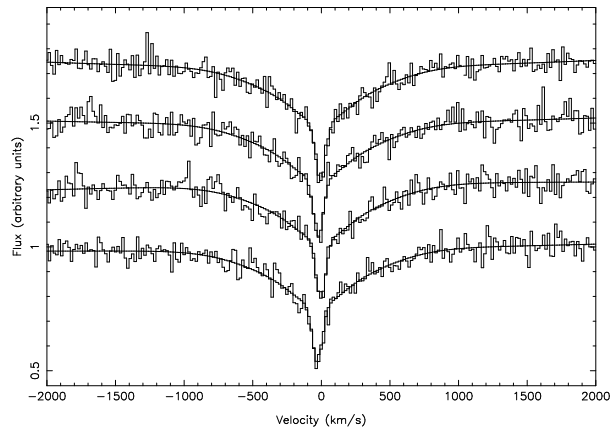} 
\caption{\label{PG1043Fig} An example of the observed spectra and multiple
Gaussian fits used to measure the radial velocities given in
Table~\ref{RVTable}. The spectra of PG\,1043+760 are shown plotted as
histograms together with the model fits shown a thick lines. The spectra are
normalized and offset by 0.25 units relative to one another. The wavelength is
given as a velocity relative to the rest wavelength of H${\alpha}$.} 
\end{figure}

 To measure the radial velocities we use a fitting process with four steps to
determine an optimum set of radial velocities. We use a least-squares fit to
one of the spectra to determine an initial shape of the model line profile. A
least squares fit of this profile to each spectrum in which the position of
the line is the only free parameter gives an initial set of radial velocities.
We use these initial radial velocities to fix the position of the H${\alpha}$
line in a simultaneous fit to all the spectra to obtain an improved model line
profile. A least squares fit of this profile to each spectrum yields the
radial velocities given in Table~\ref{RVTable}. The uncertainties quoted are
calculated by propagating the uncertainties on every data point in the spectra
right through the data reduction and analysis. These uncertainties are
reliable in most cases, but some caution must be exercised for quoted
uncertainties $\la 2$\,km\,s$^{-1}$. This corresponds to about
$1/10$ of a pixel in the original data, so systematic errors such as telluric
absorption features, uncertainties in the wavelength calibration  and motion
of the star within the slit during good seeing are certain to be a significant
source of uncertainty for these measurements. An example of the observed
spectra and multiple Gaussian fits for one star is shown in
Fig.~\ref{PG1043Fig}.

 We rebinned all the spectra onto a common wavelength scale allowing for the
measured radial velocity shifts and then formed the average spectrum of each
star shown in Fig.~\ref{AvFig}.

\begin{table*}
\caption{\label{ResultsTable}
 Summary of our radial velocity measurements for subdwarf-B stars. See
section \ref{Results} for details.}
\begin{tabular}{@{}lrrrrccccc}
Name & \multicolumn{1}{l}{N}&\multicolumn{1}{l}{$\Delta$(km\,s$^{-1}$)} &
\multicolumn{1}{c}{$\chi^2$} & $\log_{10}(p)$ & \multicolumn{1}{l}{SLY98} &
\multicolumn{1}{l}{JP98} & \multicolumn{1}{l}{AWFBL94} & UT98 &
\multicolumn{1}{l}{Notes}\\
PG\,0749+658 & 10 &$ 17.6 \pm 4.6 $& 20.24 &$ -1.78$ & $\times$ & & K5.5 & & $\star$ \\
PG\,0839+399 & 6 &$ 75.2 \pm 6.9 $& 172.21 &\mbox{\boldmath{$-34.61$}} & 2 & $\times$  \\
PG\,0849+319 & 4 &$ 44.5 \pm 4.9 $& 89.41 &\mbox{\boldmath{$ -18.53$}}& &  \\
PG\,0850+170 & 4 &$ 44.0 \pm 2.7 $& 220.23 &\mbox{\boldmath{$ < -45$}}& & $\times$ & &  \\
PG\,0907+123 & 6 &$ 66.2 \pm 5.2 $& 317.49 &\mbox{\boldmath{$ < -45$}}& & $\times$ & & \\
(PG\,0909+164) & 8 &$ 25.7 \pm 8.6 $& 14.73 &$ -1.40$  \\
PG\,0918+029 & 4 &$ 81.3 \pm 11.9 $& 114.38 &\mbox{\boldmath{$ -23.90$}}& $\times$ & & & &  \\
PG\,0919+273 & 4 &$ 25.8 \pm 5.0 $& 30.92 &\mbox{\boldmath{$ -6.05$}} & $\times$ & & & & \\
(PG\,1000+408) & 10 &$ 37.8 \pm 11.9 $& 31.57 &\mbox{\boldmath{ $-3.63$}}& & & $\times$  \\
PG\,1017$-$086& 2 &$ 61.7 \pm 9.6 $& 38.16 &\mbox{\boldmath{$ -9.19$}}&&& & \\
PG\,1018$-$047& 8 &$ 17.7 \pm 5.6 $& 13.13 &$ -1.16$&  & &&&$\star$ \\
PG\,1032+406 & 4 &$ 38.4 \pm 4.2 $& 89.89 &\mbox{\boldmath{$ -18.63$}}& &  \\
PG\,1039+219 & 6 &$ 8.4 \pm 4.2 $& 5.06 &$ -0.39$& & $\times$ &  & & $\star$\\
PG\,1040+234 & 8 &$ 5.2 \pm 3.4 $& 2.35 &$ -0.03$& & Y & K3.5 & &  $\star$ \\
PG\,1043+760 & 4 &$ 24.3 \pm 4.6 $& 30.08 &\mbox{\boldmath{$ -5.88$}}& &  \\
PG\,1047+003 & 6 &$ 13.0 \pm 6.5 $& 5.71 &$ -0.47$& &$\times$ &  & & $\star$\\
(PG\,1051+501)& 6 &$ 9.1 \pm 5.8 $& 2.33 &$ -0.10$& & $\times$ & & \\
PG\,1110+294 & 6 &$ 54.9 \pm 9.5 $& 156.96 &\mbox{\boldmath{$ -31.36$}}& \\
PG\,1114+073 & 8 &$ 6.2 \pm 2.5 $& 3.94 &$ -0.10$& $\times$ & & & &  $\star$ \\
PG\,1116+301 & 4 &$ 171.9 \pm 4.8 $&2203.13 &\mbox{\boldmath{$ < -45$}}& \\
PG\,1237+132 & 10 &$ 21.0 \pm 5.9 $& 17.06 &$ -1.32$& \\
PG\,1244+113 & 4 &$ 86.9 \pm 8.5 $& 161.99 &\mbox{\boldmath{$ -34.17$}}& & $\times$ & & \\
PG\,1248+164 & 4 &$ 114.3 \pm 3.5 $&1195.15 &\mbox{\boldmath{$ < -45$}}& \\
PG\,1300+279 & 4 &$ 95.5 \pm 3.2 $& 867.74 &\mbox{\boldmath{$ < -45$}}& \\
PG\,1303+097 & 5 &$ 9.1 \pm 4.2 $& 4.16 &$ -0.41$& \\
PG\,1329+159 & 4 &$ 27.9 \pm 3.3 $& 47.63 &\mbox{\boldmath{$ -9.59$}}& \\
PG\,1417+257 & 10 &$ 12.6 \pm 3.7 $& 17.19 &$ -1.34$& \\
(PG\,1505+074)& 4 &$ 10.9 \pm 5.3 $& 6.75 &$ -1.10$&&&& &  \\
PG\,1512+244 & 4 &$ 60.4 \pm 4.5 $& 224.53 &\mbox{\boldmath{$ < -45$}}&&&&$\times$& \\
(PG\,1553+273)& 6 &$ 8.1 \pm 2.9 $& 7.73 &$ -0.76$& \\
PG\,1616+144 & 4 &$ 3.3 \pm 5.5 $& 0.53 &$ -0.04$& \\
PG\,1619+522 & 6 &$ 43.5 \pm 5.2 $& 90.70 &\mbox{\boldmath{$ -17.32$}}& \\
PG\,1627+017 & 4 &$ 60.2 \pm 4.0 $& 273.94 &\mbox{\boldmath{$ < -45$}}&&&$\times$ & \\
(PG\,1631+267\,B) & 2 &$ 0.9 \pm 0.8 $& 0.09 &$ -0.12$&&&& & $\star$ \\
(PG\,1632+088) & 2 &$ 1.3 \pm 1.6 $& 0.16 &$ -0.16$& \\
PG\,1647+056 & 2 &$ 7.1 \pm 4.3 $& 1.90 &$ -0.78$&&&K8& \\
PG\,1653+131 & 2 &$ 3.0 \pm 3.2 $& 0.50 &$ -0.32$&&&$\times$& \\
PG\,1701+359 & 10 &$ 5.6 \pm 2.9 $& 4.75 &$ -0.07$&&&K6.5& & $\star$ \\
PG\,1710+490 & 14 &$ 14.2 \pm 5.7 $& 23.00 &$ -1.38$& \\
PG\,1716+426 & 6 &$ 131.8 \pm 3.1 $&2040.05 &\mbox{\boldmath{$ < -45$}}& 1 & && & $\star$ \\
PG\,1722+286 & 6 &$ 14.4 \pm 5.9 $& 6.29 &$ -0.55$& \\
PG\,1725+252 & 6 &$ 133.8 \pm 4.5 $&1326.41 &\mbox{\boldmath{$ < -45$}}&&&$\times$ & \\
PG\,1743+477 & 6 &$ 21.5 \pm 3.1 $& 32.10 &\mbox{\boldmath{$ -5.25$}}& \\
\end{tabular}
\end{table*}

\subsection{Criterion for variability.}
 For each star we calculate a weighted mean radial velocity. This mean is the
best estimate of the radial velocity of the star assuming this quantity is
constant. We then calculate the $\chi^2$ statistic for this ``model'', i.e.,
the goodness-of-fit of a constant to the observed radial velocities. We can
then compare the observed value of $\chi^2$ with the distribution of $\chi^2$
for the appropriate number of degrees of freedom. We then calculate the
probability of obtaining the observed value of $\chi^2$ or higher from random
fluctuations of constant value, $p$. To allow for the systematic errors
described above, we have  added a 2\,km\,s$^{-1}$ external error in quadrature
to all the radial velocity uncertainties prior to calculating these
statistics. If we find $\log_{10}(p) < -4$ we consider this to be a detection
of a binary. In our sample of 36 EHB stars, this results in a less than
0.4\,percent chance of random fluctuations producing one or more false
detections.

\subsection{\label{Results}Results}
The observed values of $\chi^2$ and $\log_{10}(p)$ and the number of measured
radial velocities, N, are given for all the targets in our sample in
Table~\ref{ResultsTable}.  Stars which were observed but are not EHB stars are
shown in parentheses. Stars which we consider to be binaries are denoted by
displaying $\log_{10}(p)$ in bold type. In column 3 we give the maximum
difference between the observed radial velocities, $\Delta$.    In column 6
(SLY98) we note whether Saffer, Livio \& Yungelson (1998) saw a marginal
detection (2) or a positive detection (1) of a radial velocity shift or failed
to detect any radial velocity shift ($\times$). In column 7 (JP98) we note
whether Jeffery \& Pollacco (1998) saw spectral features due to a cool
companion (Y) or failed to detect a companion ($\times$). In column 8
(AWFBL98) we note whether the BVRI photometry of Allard et~al. (1994) failed
to detect a companion ($\times$) or note the spectral type of the companion if
it was detected.  In column 8 (UT98) we note stars for which Ulla \& Thejll
did not detect any infrared excess due to a companion from their JHK
photometry ($\times$). Stars for which comments can be found  in section
\ref{Notes} are noted in column 9. There are 36 EHB stars in our sample, 21 of
which are binaries. With the  exception of PG\,1716+426 and, perhaps,
PG\,0839+399, these are all new detections.

\section{ Notes on individual objects.}\label{Notes}
 In this section we note previous results for our targets and any other
remarkable or peculiar characteristics. 

\begin{figure*}
\psfig{file=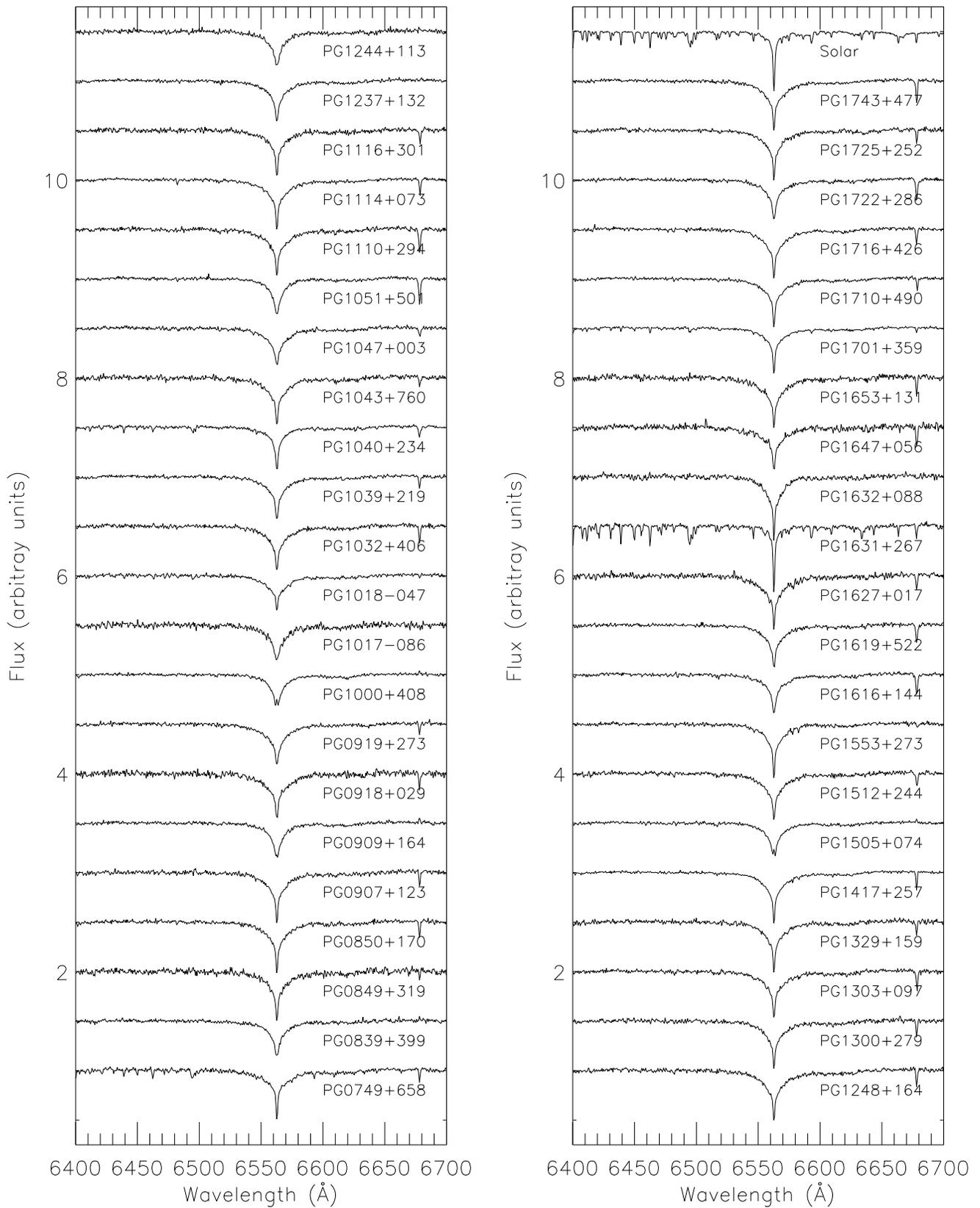,height=0.95\textheight} 
\caption{\label{AvFig} The average spectrum of each star. The spectra are
normalized and offset by 0.5 units relative to one another. A spectrum of the
twilight sky labelled ``Solar'' is shown for comparison. }
\end{figure*}

\begin{description}
\item[\bf PG\,0749+658]{ Late-type spectral features can be
seen in the average spectrum of this star shown in Fig.~\ref{AvFig}. }
\item[\bf PG\,1018$-$047]{There are weak spectral features due to a late-type
companion visible in our spectra.} 
\item[\bf PG\,1039+219]{ This star in listed in Jeffery \& Pollacco as
Ton\,1273.} 
\item[\bf PG\,1040+234]{ Spectral features due to the companion are seen in
our blue spectra of this star, notably the G-band and Ca\,II~H\&K spectral
lines, and some weak features can also be seen around H${\alpha}$
(Fig.~\ref{AvFig}). } 
\item[\bf PG\,1047+003]{This is a pulsating sdB variable star (Bill\`{e}res
et~al. 1997).} 
\item[\bf PG\,1114+073]{ Saffer, Livio \& Yungelson  list this star as 
PG\,1114+072.} 
%
%
\item[\bf PG\,1631+267]{This star has a bright G-type companion which
dominates the spectrum around the H${\alpha}$ line so the radial velocities
quoted here refer to the G-star companion to the sdB star, which we denote
PG\,1631+267\,B. } 
\item[\bf PG\,1701+359]{ Theissen et~al. noted spectral features from a cool
star in their spectra. Spectral features from a cool star are also visible in
our spectra around the H$\alpha$ line. } 
\item[\bf PG\,1716+426]{ Geffert (1998) considers the galactic orbit of this
star based on the HIPPARCOS astrometry and a radial velocity measurement of
$-10.6\pm30$\,km\,s$^{-1}$. Clearly, this calculation needs to be revised.} 
\end{description}

\section{Discussion}
\subsection{Estimating the binary fraction.}

 The probability of detecting $N_B$ binaries in a sample of $N$ stars which
have a binary fraction of $f$ is 
\[\frac{N!}{(N-N_B)!N_B!}(\bar{d}f)^{N_B}(1-\bar{d}f)^{N-N_B}\]
where $\bar{d}$ is the fraction of all binaries detected by the survey
averaged over all orbital periods. For our survey, $N_B=21$ and $N=36$. We can
set a lower limit to $f$ by assuming $\bar{d}=1$, i.e., the lower limit to $f$
is set by assuming we have detected all the binaries in our sample. In this
case we expect that the lower limit to the binary fraction will be about
21/33=58.3\,percent. In fact, the distribution of $f$ calculated with the
expression above with $\bar{d}=1$ is approximately Gaussian with a maximum at
$f=0.60$ and a standard deviation of 0.08, i.e., the absolute lower limit to
$f$ from our survey is 60$\pm$8\,percent. 

\begin{figure}
\psfig{file=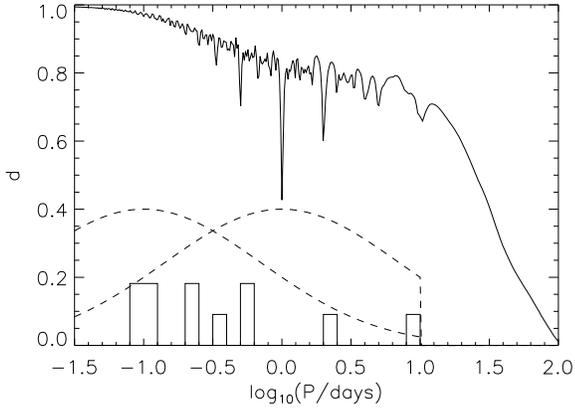,width=0.48\textwidth} 
\caption{\label{BfracFig} The fraction of binaries detected by our survey,
$d$, as a function of the orbital period, $P$ (thick line). The Gaussian
weighting functions described in the text  (dashed lines) and the distribution
of orbital periods for known EHB binaries given in Table~\ref{PerTable}
(histogram) are also shown. } 
\end{figure}

 We calculated the fraction of all binaries of a given orbital period, $P$,
detected by our survey, $d$, as follows. We assume that the EHB star and its
companion both have a mass of 0.5\Msolar. We can then calculate the orbital
speed of the EHB star, $V_{\rm orb}$,  assuming a circular orbit. We assume
the that orbits are circular because a common envelope phase will quickly
reduce the eccentricity of an orbit and no post-common envelope systems are
observed to have any appreciable eccentricity. For a given star for which we
have $N_{\rm obs}$ radial velocity measurements we can then use the actual
dates of observation, $T_j, j=1\dots N_{\rm obs}$  to calculate radial
velocities for a hypothetical binary with an edge-on orbit from $V_{\rm
orb}\sin(\phi_i)$, where $\phi = (T_j-T_0)/P$. These values are used to
calculate the value of chi-squared for this hypothetical binary, $\chi^2_{\rm
max}$, using the actual radial velocity uncertainties for the $N_{\rm obs}$
observations given in Table~\ref{RVTable} including 2\,km\,s$^{-1}$ additional
systematic uncertainty. The calculation is repeated for 50 values of the $T_0$
and the average value of $\chi^2_{\rm max}$ is taken. We can then compare the
value of $\chi^2_{\rm max}$ for this hypothetical binary to the value of
chi-squared required to exactly satisfy our detection criterion, $\chi^2_{\rm
crit}$. If $\chi^2_{\rm max} < \chi^2_{\rm crit}$ then no binaries with that
orbital period, mass and eccentricity  will be detected by our observations.
\footnote{This is not strictly true. Adding random fluctuations to the $V_{\rm
orb}\sin(\phi_i)$ values can result in detections in cases where $\chi^2_{\rm
max}$ is only slightly less then $\chi^2_{\rm crit}$. Similarly, noise can
prevent some detections when $\chi^2_{\rm max}$ is slightly greater than
$\chi^2_{\rm crit}$. The overall effect is negligible when the detection
effeciency is averaged over a wide range of  orbital periods as we have done.}
 Otherwise, we can calculate the projected orbital velocity for which
$\chi^2_{\rm max} = \chi^2_{\rm crit}$, $K_{\rm crit} = V_{\rm orb} \sin i$ for
some orbital inclination $i$. For randomly oriented orbits, $i$ is distributed
as $\cos i$ so the fraction of binaries detected for this combination of
observations, period, mass etc.  is simply $d=\sqrt{1-(K_{\rm
crit}/V_{\rm orb})^2}$. We have calculated this detection efficiency for
20\,000 orbital periods distributed uniformly in $\log_{10}(P)$ over the
range $ -1.5 \le \log_{10}(P/\rm d) \le 2$ for every EHB star we observed and
used these values to calculate the average detection efficiency for stars in
our sample, $d$. The results are shown in Fig.~\ref{BfracFig}, where the value
of $d$ has been binned into 400 groups of 50 periods.

To calculate the binary fraction of EHB stars, we need to know $\bar{d}$, the
weighted mean of $d$ over the period distribution of EHB binaries.
Unfortunately, the period distribution of EHB binaries is very poorly known.
The existing observational data for EHB stars with measured orbital periods is
rather scarce, but is summarised in Table~\ref{PerTable} and is also shown in
Fig.~\ref{BfracFig}.  We are not aware of any reliable predictions for the
orbital period distribution based on models of the evolution of EHB stars.
From the size of the radial velocity shifts given in Table~\ref{ResultsTable}
we can set an upper limit to the orbital period of the binaries we have found.
These are typically tens-of-days, so the actual orbital periods are likely to
be $\la 10$\,d.\footnote{Observations to determine the actual orbital periods
are being undertaken at the time of writing.}

 In the absence of any good determination of the orbital period distribution
of binary EHB stars for longer orbital periods, we consider the binary
fraction for  short orbital period binaries only ($P\la 10$\,d).  We can see
from Fig.~\ref{BfracFig} that  any reasonable period distribution will give a
mean detection efficiency over $ -1.5 \le \log_{10}(P/\rm d) \le 1$ of about
85\,percent. For example, the unweighted average over this range of
$\log_{10}(P)$ is 86.6\,percent, which implies a binary fraction of
69$\pm$9\,percent. If, for the sake of argument, we assume a distribution  for
$\log_{10}(P)$ which is a Gaussian function with a mean of $\log_{10}(P/\rm
d)=0$ and a full-width at half-maximum of $\log_{10}(P/\rm d)=2$, the mean
detection efficiency over  $ -1.5 \le \log_{10}(P/\rm d) \le 1$ is
84.7\,precent so we obtain a binary fraction of 70$\pm$9\,percent. If we
change the mean of the Gaussian function to $\log_{10}(P/\rm d)=-1$, the mean
detection efficiency over the same range of  $\log_{10}(P)$ is 91.1\,percent
and the binary fraction  is 65$\pm$9\,percent. These are all {\it ad-hoc}
assumptions for the period distribution of binary EHB stars, but they do show
that the fraction of EHB stars which are short-period binaries is about 2/3
for any reasonable period distribution. 

 This calculation is also insensitive to the assumed mass ratios of the
binaries. If we assume the companions have a mass of 1\Msolar, the
lower limits to the binary fraction we derive are reduced by 1--2\,percent.
The mean companion mass is unlikely to be larger than 1\Msolar\ because a
main-sequence or sub-giant star of this mass would be easily visible in the
spectrum and the upper limit to the white dwarf companion mass is, of course,
the Chandrasekhar mass of 1.4\Msolar. We do not expect there to be large
numbers of neutron star or black hole companions to EHB stars. If the
companions have a lower mean mass, our detection efficiency would be
lower than the value calculated, so the the minimum binary fraction we would
derive would be higher. In summary, the minimum binary fraction implied by our
observations is about 69$\pm$9\,percent and this result is not strongly
dependent on the assumed distributions of period or mass ratios for short
period EHB binaries.

Of course, if there are also binary EHB stars with longer periods, these would
not be detected as frequently by our survey as the shorter period binaries, so
the binary fraction may be much higher than 2/3. At some point the orbital
period is too long for the binary to be relevant to this discussion. The
binaries of interest are those for which the orbital separation now is less
than the size of a red giant star near the tip of the red giant branch (RGB).
In these cases, we can say that the companion has influenced the formation
process of the EHB star. The radii of red giants near the tip of the RGB are
$\approx 100\Rsolar$, which corresponds to orbital periods of a hundred days
or more. We can see from Fig.~\ref{BfracFig} that most binary EHB stars with
orbital periods of tens-of-days would be missed by our survey. Therefore, if
about 1/3 of EHB stars are binaries with orbital periods 10\,d$\la P \la
$100\,d, then our results are consistent with  all EHB stars being formed
through interactions with a companion star.

\begin{table}
\caption{\label{PerTable}
 Measured orbital periods and companion masses, $M_2$
for binary EHB stars. The lower limits to the companion masses have
been calculated from the projected orbital velocity assuming a mass for
the EHB star of 0.5\Msolar. White dwarf companions are denoted ``WD'',
otherwise the spectral type of the companionm if known, is given.}
\begin{tabular}{@{}lrrll}
Name & \multicolumn{1}{l}{Period}&\multicolumn{1}{l}{$M_2$} & 
\multicolumn{1}{l}{Companion } & Ref.\\
& \multicolumn{1}{l}{(days)}&\multicolumn{1}{l}{(\Msolar)} & 
\multicolumn{1}{l}{type } &     \\
 KPD\,0422+5421 &  0.090  &   0.53  & WD & 2 \\
 KPD\,1930+2752 &  0.095  & 0.97    & WD & 7 \\
 PG\,1336$-$018 &  0.101  &  0.15   & M5 & 5 \\
 HW~Vir         &  0.117  &  0.14   & dM & 4 \\
 PG\,1432+159   &  0.225  & $>$0.29 & WD & 1 \\
 PG\,2345+318   &  0.241  & $>$0.38 & WD & 1 \\
 PG\,1101+249   &  0.354  & $>$0.42 & WD & 1 \\
 PG\,0101+039   &  0.570  & $>$0.37 & WD & 1 \\
 PG\,1247+553   &  0.599  & $>$0.09 & -- & 3 \\
 PG\,1538+269   &  2.501  & 0.6     & WD & 6 \\
 PG\,0940+068   &  8.33   & $>$0.63 & -- & 3 \\
\hline
\end{tabular} \\
1.~Moran et~al. (1999); 2.~Orosz \& Wade (1999); 3.~Maxted et~al.
(2000); 4.~Wood \& Saffer (1999); 5.~Kilkenny et~al. (1998); 6.~Ritter \& Kolb
(1998); 7.~Maxted, Marsh \& North (2000). 
\end{table}

\subsection{Other surveys.}
 We note that there is no significant radial velocity shift in any of the
stars for which there is evidence of a cool companion (PG\,0749+658,
PG\,1018$-$047,  PG\,1040+234, PG\,1647+056 and PG\,1701+359). There is a bias
in our sample in the sense that the orbital separation of a binary with two
0.5\Msolar\ stars may be too small to contain a main-sequence or sub-giant
K-star for the shorter orbital periods  where our survey has the greatest
sensitivity. However, our observations are still quite sensitive to periods of
a day or more, at least for PG\,1040+234 and PG\,1701+359. The orbital
separation for an orbital period of a few days is several solar radii. This
may suggest that there is a real trend for EHB stars with K-type companions to
have long orbital periods. 

 Of the six stars observed by SLY which also appear in our sample, we find
four to be binaries. Two of these binaries were not detected by SLY
(PG\,0918+029 and PG\,0919+273), one was noted as a marginal detection
(PG\,0839+399) and one as positive detection (PG\,1716+426). The shifts seen
by us for PG\,0918+029 suggest that it is at the limit of detection for the
method employed by SLY. Curiously, the shift we observed for PG\,0839+399, for
which SLY note a marginal detection, is smaller than that of PG\,0918+029,
which suggests that our observations have not yet sampled the  full range of
radial velocity shifts for this star. The shift of only 25.8\,km\,s$^{-1}$ we
measured for PG\,0919+273 would not have been seen by SLY. 

\subsection{Post-EHB stars.}
 Of the stars observed which we have excluded from this discussion, 5 appear
to be stars which have evolved away from the extreme horizontal branch. One of
these post-EHB stars a good candidate to be a binary from our data
(PG\,1000+408). Observations by Green (2000) have shown that PG\,1000+408 is
indeed a binary with an orbital period near 1\,day. There are too few stars in
this sub-sample to derive useful limits on the binary fraction of these stars,
though this would obviously be an interesting number because we would expect
it to be similar to the binary fraction for normal EHB stars if the two
groups of stars are related as we have suggested.

\subsection{Selection effects.}
 One advantage of choosing objects from the PG survey is that we were able to
choose brighter stars based on their photographic magnitudes without
introducing a bias in our sample towards short period binaries. This is
because the majority of the short period binaries have white dwarf or K/M
dwarf companions, both of which contribute a negligible amount of light in the
blue region of the spectrum on which the PG survey is based. One type of
binary we are biased against are those containing brighter F/G-type
companions. Most of these binaries were excluded from the PG survey because
stars showing the Ca\,II~K line were assumed to be main-sequence subdwarfs
with normal colours which appeared bluer due to the substantial uncertainty in
the photographic photometry on which survey is based (Green, Schmidt \&
Liebert 1986). In fact, a substantial fraction of these stars may be sdB stars
with F/G-type companions (Kilkenny et~a., 1997). Some sdB stars with F/G-type
companion were are included in the PG survey, e.g., PG\,1631+267. Although we
cannot measure the radial velocities of the sdB star from the H$\alpha$ line
in these cases, we can measure the radial velocities of the F/G star from its
H$\alpha$ line, if it dominates at the wavelength, or from the many other
absorption lines should the H$\alpha$ line be a blend of the sdB star and the
F/G star  H$\alpha$ line. In fact, these radial velocities are preferable to
the sdB velocities in some ways as they are more accurate and we can estimate
the mass of the companion star from its spectral type.

\subsection{Triple stars}
 The effect of the selection criterion against F/G companion stars applied 
to the PG survey is difficult to judge without knowing the actual fraction of
sdB stars with F/G-type companions that have been ``lost'' from the survey. We
note that this effect also biases the results of Allard et~al. (1994),
Ferguson, Green \& Liebert (1984) and Jeffery \& Pollacco (1998).
Nevertheless, these authors still find 1/2\,--\,2/3 of their sample have cool
companion stars. We have already noted that the EHB stars we have observed to
be binaries do not have cool companions. The total binary fraction is then 2/3
short period sdB stars without cool companions plus 1/2\,--\,2/3 with cool
companions in the PG survey plus the ``lost binaries'' with cool companions
excluded from the PG survey  minus a small fraction of short period sdB stars
with cool companions. This number is clearly greater than 1, an apparent
paradox which is easily explained by some of the sdB stars being triple stars,
i.e., short period sdB stars with M-dwarf or white dwarf companions and a
distant F/G-type cool companion which was not involved with the evolution of
the inner pair.

\section{Conclusion}

 We have measured 205  precise radial velocities for 36 extreme horizontal
branch  stars to look for variability due to a close binary companion. We
found that 21 of our stars are positively identified as short-period binaries.
All but one or two of these are new identifications. We conclude that at least
2/3 of all EHB stars are short-period binaries. The orbital separations of
these binaries is much less than the size of the star during the red giant
phase which almost certainly preceded its emergence as an EHB star. We
conclude that some kind of interaction with a binary companion, perhaps in a
common envelope phase, is fundamental to the formation process for the
majority of EHB stars.

\section*{Acknowledgements}
 PFLM was supported by a PPARC post-doctoral grant.  The Isaac Newton
Telescope is operated on the island of La Palma by the Isaac Newton Group in
the Spanish Observatorio del Roque de los Muchachos of the Instituto de
Astrofisica de Canarias. PFLM would like to thank E.M. Green and R.A. Saffer
for many informative discussions on the subject of sdB stars. We would like to
thank Rex Saffer for providing us with his results for sdB stars.
We would like to thank Martin Altmann for his spectrum of PG\,0907+123.

%
\begin{table}
\caption{\label{RVTable} Measured heliocentric radial velocities.}
\begin{tabular}{@{}lrr}
Name & \multicolumn{1}{l}{HJD} & \multicolumn{1}{l}{Radial velocity}  \\
& \multicolumn{1}{l}{$-$2451600} & (km\,s$^{-1}$) \\
PG\,0749+658 \\
&    46.3402&    $-$8.7 $\pm$   3.9\\
&    46.3412&   $-$10.8 $\pm$   4.3\\
&    51.3388&   $-$11.9 $\pm$   2.4\\
&    51.3405&   $-$10.4 $\pm$   2.2\\
&    54.3339&     5.7 $\pm$   3.9\\
&    54.3356&    $-$0.7 $\pm$   3.3\\
&    56.3462&   $-$10.9 $\pm$   2.0\\
&    56.3478&   $-$11.5 $\pm$   2.1\\
&    57.3379&    $-$4.0 $\pm$   2.8\\
&    57.3396&    $-$6.1 $\pm$   2.4\\
PG\,0839+399 \\
&    46.3498&    52.4 $\pm$   8.6\\
&    46.3562&    31.9 $\pm$   8.8\\
&    51.3528&    64.7 $\pm$   4.6\\
&    51.3677&    55.8 $\pm$   4.3\\
&    54.3695&    $-$5.3 $\pm$   5.1\\
&    54.3820&    $-$7.0 $\pm$   4.8\\
PG\,0849+319 \\
&    46.3657&    93.0 $\pm$   4.3\\
&    46.3733&    82.0 $\pm$   4.3\\
&    51.3816&   119.3 $\pm$   4.1\\
&    51.3891&   127.6 $\pm$   4.1\\
PG\,0850+170 \\
&    46.4208&    69.2 $\pm$   1.8\\
&    46.4333&    66.3 $\pm$   1.9\\
&    54.3966&    27.0 $\pm$   2.2\\
&    54.4067&    25.2 $\pm$   2.0\\
PG\,0907+123 \\
&    47.4116&    87.3 $\pm$   2.7\\
&    47.4323&    83.2 $\pm$   2.2\\
&    53.3980&   101.1 $\pm$   4.0\\
&    53.4084&    92.6 $\pm$   6.7\\
&    54.4213&    34.9 $\pm$   3.4\\
&    54.4351&    37.8 $\pm$   1.9\\
PG\,0909+164 \\
&    47.4525&    59.4 $\pm$   6.5\\
&    47.4677&    42.6 $\pm$   6.9\\
&    53.3753&    43.4 $\pm$   5.4\\
&    53.3856&    52.0 $\pm$   5.5\\
&    56.3918&    51.4 $\pm$   5.5\\
&    56.4021&    48.8 $\pm$   5.4\\
&    57.3921&    68.3 $\pm$   5.1\\
&    57.4024&    57.4 $\pm$   5.2\\
PG\,0918+029 \\
&    47.4808&    35.0 $\pm$   5.9\\
&    47.4861&    26.1 $\pm$  11.0\\
&    53.3633&    95.0 $\pm$   5.0\\
&    53.3669&   107.4 $\pm$   4.5\\
PG\,0919+273 \\
&    53.3518&   $-$77.2 $\pm$   4.1\\
&    53.3562&   $-$83.3 $\pm$   3.8\\
&    54.4461&   $-$57.5 $\pm$   3.2\\
&    54.4505&   $-$58.3 $\pm$   3.2\\
\\
\\
\\
\\
\\
\\
\\
\\
\\
\\
\end{tabular}
\end{table}
\begin{table}
\contcaption{}
\begin{tabular}{@{}lrr}
Name & \multicolumn{1}{l}{HJD} & \multicolumn{1}{l}{Radial velocity}  \\
& \multicolumn{1}{l}{$-$2451600} & (km\,s$^{-1}$) \\
PG\,1000+408 \\
&    49.3644&    66.4 $\pm$  11.4\\
&    49.3687&    76.4 $\pm$   9.4\\
&    51.3991&   102.6 $\pm$   4.2\\
&    51.4054&    95.6 $\pm$   4.1\\
&    51.4197&    91.7 $\pm$   4.5\\
&    51.4260&    83.9 $\pm$   4.3\\
&    56.3593&    85.4 $\pm$   4.9\\
&    56.3661&   102.0 $\pm$   4.9\\
&    57.3492&    77.8 $\pm$   4.4\\
&    57.3575&    84.7 $\pm$   3.8\\
PG\,1017$-$086 \\
&    46.4502&   $-$66.2 $\pm$   7.1\\
&    46.4614&    $-$4.5 $\pm$   6.4\\
PG\,1018$-$047 \\
&    46.4713&    33.5 $\pm$   3.8\\
&    46.4765&    21.7 $\pm$   3.7\\
&    54.4607&    24.8 $\pm$   3.2\\
&    54.4678&    33.0 $\pm$   3.2\\
&    56.4141&    27.6 $\pm$   4.0\\
&    56.4192&    29.1 $\pm$   4.0\\
&    57.4141&    15.8 $\pm$   4.1\\
&    57.4192&    24.5 $\pm$   4.0\\
PG\,1032+406 \\
&    49.3760&    $-$8.6 $\pm$   4.2\\
&    49.3777&   $-$15.3 $\pm$   3.4\\
&    51.4124&    23.1 $\pm$   2.4\\
&    51.4141&    22.5 $\pm$   2.4\\
PG\,1039+219 \\
&    53.4438&     0.4 $\pm$   2.9\\
&    53.4500&    $-$2.5 $\pm$   2.9\\
&    56.4266&    $-$7.5 $\pm$   3.3\\
&    56.4330&    $-$7.6 $\pm$   3.0\\
&    57.4266&    $-$3.5 $\pm$   3.0\\
&    57.4328&     0.8 $\pm$   2.9\\
PG\,1040+234 \\
&    53.4582&    11.5 $\pm$   2.6\\
&    53.4641&     8.1 $\pm$   2.8\\
&    54.4761&     9.7 $\pm$   2.2\\
&    54.4821&     8.4 $\pm$   2.2\\
&    56.4411&    11.0 $\pm$   2.5\\
&    56.4471&    12.6 $\pm$   2.4\\
&    57.4406&     7.4 $\pm$   2.4\\
&    57.4466&     9.1 $\pm$   2.4\\
PG\,1043+760 \\
&    49.3849&   $-$28.1 $\pm$   3.2\\
&    49.3959&    $-$3.8 $\pm$   3.3\\
&    51.4332&   $-$21.7 $\pm$   3.1\\
&    51.4402&    $-$5.8 $\pm$   3.1\\
PG\,1047+003 \\
&    53.4284&    $-$3.2 $\pm$   5.0\\
&    53.4357&    $-$0.0 $\pm$   5.1\\
&    56.4560&    $-$1.7 $\pm$   5.2\\
&    56.4617&   $-$11.1 $\pm$   5.1\\
&    57.4579&    $-$9.8 $\pm$   3.8\\
&    57.4690&   $-$12.9 $\pm$   3.8\\
\\
\\
\\
\\
\\
\\
\\
\\
\end{tabular}
\end{table}
\begin{table}
\contcaption{}
\begin{tabular}{@{}lrr}
Name & \multicolumn{1}{l}{HJD} & \multicolumn{1}{l}{Radial velocity}  \\
& \multicolumn{1}{l}{$-$2451600} & (km\,s$^{-1}$) \\
PG\,1051+501 \\
&    53.5281&  $-$133.8 $\pm$   4.7\\
&    53.5355&  $-$127.5 $\pm$   5.0\\
&    56.3752&  $-$126.8 $\pm$   5.8\\
&    56.3811&  $-$128.9 $\pm$   5.6\\
&    57.3692&  $-$124.7 $\pm$   3.4\\
&    57.3796&  $-$130.3 $\pm$   3.3\\
PG\,1110+294 \\
&    53.4747&    11.1 $\pm$  11.9\\
&    53.4845&    14.1 $\pm$   9.1\\
&    56.5069&    $-$0.6 $\pm$   2.6\\
&    56.5167&    $-$0.3 $\pm$   2.6\\
&    57.4808&   $-$39.3 $\pm$   2.8\\
&    57.4906&   $-$40.8 $\pm$   2.8\\
PG\,1114+073 \\
&    47.5230&     8.6 $\pm$   2.6\\
&    47.5338&     7.7 $\pm$   2.4\\
&    54.5177&     5.7 $\pm$   1.5\\
&    54.5259&    11.3 $\pm$   1.5\\
&    56.4693&    12.0 $\pm$   2.0\\
&    56.4749&     8.5 $\pm$   1.9\\
&    57.5014&     9.5 $\pm$   2.0\\
&    57.5070&     8.6 $\pm$   2.0\\
PG\,1116+301 \\
&    53.5014&   $-$89.4 $\pm$   3.8\\
&    53.5164&   $-$86.4 $\pm$   2.7\\
&    56.5285&    82.5 $\pm$   2.8\\
&    56.5386&    79.5 $\pm$   2.7\\
PG\,1237+132 \\
&    46.4896&   $-$36.1 $\pm$   4.8\\
&    46.5032&   $-$31.6 $\pm$   4.8\\
&    54.5637&   $-$31.3 $\pm$   4.4\\
&    54.5772&   $-$36.4 $\pm$   4.2\\
&    55.6295&   $-$33.5 $\pm$   3.9\\
&    55.6431&   $-$35.5 $\pm$   3.9\\
&    56.5768&   $-$29.1 $\pm$   4.2\\
&    56.5904&   $-$34.4 $\pm$   4.1\\
&    57.5443&   $-$50.1 $\pm$   4.2\\
&    57.5578&   $-$44.5 $\pm$   4.1\\
PG\,1244+113 \\
&    46.5409&    61.9 $\pm$   5.8\\
&    46.5505&    66.3 $\pm$   6.1\\
&    54.5907&   $-$14.0 $\pm$   6.9\\
&    54.6003&   $-$21.9 $\pm$   5.9\\
PG\,1248+164 \\
&    46.5672&    40.6 $\pm$   1.8\\
&    46.5875&    44.5 $\pm$   1.8\\
&    56.5529&   $-$63.8 $\pm$   2.9\\
&    56.5632&   $-$69.8 $\pm$   3.0\\
PG\,1300+279 \\
&    46.6051&    52.0 $\pm$   2.2\\
&    46.6166&    49.0 $\pm$   2.2\\
&    56.6042&   $-$34.2 $\pm$   2.5\\
&    56.6156&   $-$43.4 $\pm$   2.3\\
PG\,1303+097 \\
&    46.6356&    28.9 $\pm$   2.3\\
&    46.6548&    29.3 $\pm$   2.6\\
&    56.6651&    36.1 $\pm$   3.4\\
&    57.5763&    32.9 $\pm$   2.5\\
&    57.5954&    27.0 $\pm$   2.5\\
\end{tabular}
\end{table}
\begin{table}
\contcaption{}
\begin{tabular}{@{}lrr}
Name & \multicolumn{1}{l}{HJD} & \multicolumn{1}{l}{Radial velocity}  \\
& \multicolumn{1}{l}{$-$2451600} & (km\,s$^{-1}$) \\
PG\,1329+159 \\
&    46.6683&    $-$1.2 $\pm$   2.8\\
&    46.6720&   $-$11.6 $\pm$   2.7\\
&    56.6277&    16.2 $\pm$   1.9\\
&    56.6351&    10.3 $\pm$   2.0\\
PG\,1417+257 \\
&    46.6820&     2.0 $\pm$   1.5\\
&    46.6945&     1.6 $\pm$   1.4\\
&    51.5643&    $-$7.3 $\pm$   2.6\\
&    51.5708&    $-$9.4 $\pm$   2.6\\
&    55.6587&    $-$1.6 $\pm$   1.4\\
&    55.6713&    $-$2.8 $\pm$   1.4\\
&    56.6969&    $-$5.0 $\pm$   1.6\\
&    56.7094&    $-$1.2 $\pm$   1.7\\
&    57.6094&     3.2 $\pm$   2.5\\
&    57.6158&     0.7 $\pm$   2.5\\
PG\,1505+074 \\
&    53.5699&     7.4 $\pm$   4.9\\
&    53.5727&     3.9 $\pm$   4.7\\
&    57.6275&    $-$3.5 $\pm$   1.9\\
&    57.6380&     5.6 $\pm$   1.9\\
PG\,1512+244 \\
&    53.5782&  $-$101.6 $\pm$   3.4\\
&    53.5837&   $-$94.8 $\pm$   3.2\\
&    57.6469&   $-$43.8 $\pm$   2.9\\
&    57.6525&   $-$41.2 $\pm$   3.0\\
PG\,1553+273 \\
&    53.5917&    71.8 $\pm$   2.0\\
&    53.5987&    71.4 $\pm$   2.1\\
&    56.6439&    77.0 $\pm$   1.9\\
&    56.6509&    79.5 $\pm$   2.0\\
&    57.6599&    79.0 $\pm$   1.8\\
&    57.6670&    75.1 $\pm$   1.7\\
PG\,1616+144 \\
&    53.6083&   $-$50.0 $\pm$   4.9\\
&    53.6160&   $-$50.0 $\pm$   4.8\\
&    57.6805&   $-$46.7 $\pm$   2.4\\
&    57.6958&   $-$47.0 $\pm$   2.3\\
PG\,1619+522 \\
&    46.7032&   $-$70.1 $\pm$   3.4\\
&    46.7081&   $-$75.4 $\pm$   3.3\\
&    51.5444&   $-$67.4 $\pm$   3.3\\
&    51.5518&   $-$68.0 $\pm$   3.4\\
&    53.6887&   $-$31.9 $\pm$   4.0\\
&    53.6937&   $-$40.1 $\pm$   3.9\\
PG\,1627+017 \\
&    53.6236&  $-$125.0 $\pm$   3.3\\
&    53.6261&  $-$126.9 $\pm$   3.2\\
&    54.6144&   $-$66.7 $\pm$   2.5\\
&    54.6169&   $-$69.0 $\pm$   2.5\\
PG\,1631+267 \\
&    53.6559&   $-$42.1 $\pm$   0.6\\
&    53.6677&   $-$41.3 $\pm$   0.6\\
PG\,1632+088 \\
&    53.6329&   189.6 $\pm$   1.2\\
&    53.6398&   191.0 $\pm$   1.1\\
PG\,1647+056 \\
&    54.6305&  $-$106.3 $\pm$   3.0\\
&    54.6521&  $-$113.4 $\pm$   3.1\\
PG\,1653+131 \\
&    54.6724&     4.3 $\pm$   2.3\\
&    54.6872&     7.4 $\pm$   2.3\\
\end{tabular}
\end{table}
\begin{table}
\contcaption{}
\begin{tabular}{@{}lrr}
Name & \multicolumn{1}{l}{HJD} & \multicolumn{1}{l}{Radial velocity}  \\
& \multicolumn{1}{l}{$-$2451600} & (km\,s$^{-1}$) \\
PG\,1701+359 \\
&    46.7176&  $-$120.7 $\pm$   1.5\\
&    46.7259&  $-$121.2 $\pm$   1.5\\
&    51.5771&  $-$117.2 $\pm$   2.5\\
&    51.5822&  $-$117.6 $\pm$   2.3\\
&    54.7222&  $-$119.4 $\pm$   1.5\\
&    54.7304&  $-$122.8 $\pm$   1.5\\
&    55.7250&  $-$118.4 $\pm$   1.4\\
&    55.7333&  $-$118.7 $\pm$   1.4\\
&    56.7198&  $-$121.9 $\pm$   1.6\\
&    56.7281&  $-$118.3 $\pm$   1.6\\
PG\,1710+490 \\
&    46.7328&   $-$56.6 $\pm$   3.4\\
&    46.7354&   $-$59.3 $\pm$   3.4\\
&    51.5869&   $-$52.8 $\pm$   4.3\\
&    51.5885&   $-$51.8 $\pm$   4.3\\
&    51.6170&   $-$49.4 $\pm$   2.5\\
&    51.6208&   $-$49.4 $\pm$   2.3\\
&    53.7003&   $-$49.3 $\pm$   2.5\\
&    53.7040&   $-$54.5 $\pm$   2.5\\
&    54.7094&   $-$60.4 $\pm$   2.2\\
&    54.7131&   $-$58.9 $\pm$   2.2\\
&    55.7121&   $-$55.8 $\pm$   2.3\\
&    55.7158&   $-$61.7 $\pm$   2.4\\
&    56.7569&   $-$54.9 $\pm$   2.6\\
&    56.7618&   $-$47.4 $\pm$   5.2\\
PG\,1716+426 \\
&    46.7446&   $-$52.4 $\pm$   2.1\\
&    46.7552&   $-$53.5 $\pm$   2.2\\
&    51.5972&    58.3 $\pm$   2.3\\
&    51.6078&    48.2 $\pm$   2.4\\
&    57.7333&   $-$73.5 $\pm$   2.0\\
&    57.7481&   $-$71.5 $\pm$   1.9\\
PG\,1722+286 \\
&    51.7498&   $-$47.5 $\pm$   4.1\\
&    51.7551&   $-$38.9 $\pm$   4.4\\
&    53.7329&   $-$33.2 $\pm$   4.3\\
&    53.7382&   $-$40.3 $\pm$   4.2\\
&    57.7091&   $-$37.2 $\pm$   2.6\\
&    57.7195&   $-$35.6 $\pm$   2.7\\
PG\,1725+252 \\
&    51.7613&   $-$63.5 $\pm$   6.1\\
&    51.7629&   $-$67.0 $\pm$   9.1\\
&    53.7578&    38.0 $\pm$   3.5\\
&    53.7610&    40.2 $\pm$   4.2\\
&    54.7494&   $-$93.7 $\pm$   1.6\\
&    54.7554&   $-$86.7 $\pm$   1.8\\
PG\,1743+477 \\
&    53.7123&    38.3 $\pm$   2.1\\
&    53.7220&    28.7 $\pm$   2.2\\
&    55.6932&    39.3 $\pm$   1.8\\
&    55.7029&    45.3 $\pm$   1.9\\
&    56.7376&    44.7 $\pm$   2.1\\
&    56.7474&    50.2 $\pm$   2.2\\
\end{tabular}
\end{table}

\label{lastpage}

\end{document}